\documentclass[11pt]{article}
\usepackage{amsmath,amssymb,amsthm}
\usepackage[textwidth=18cm,textheight=22cm]{geometry}
\usepackage{graphicx}
\usepackage{psfrag}
\usepackage{array}

\renewcommand{\d}{\mathrm{d}}
\newcommand{\bt}{\boldsymbol{t}}
\newcommand{\bu}{\boldsymbol{u}}

\newcommand{\be}{\begin{equation}}
\newcommand{\ee}{\end{equation}}
\begin{document}

\title{\sc Genus-zero Whitham hierarchies \\in  conformal-map dynamics \thanks{Partially supported by MEC
project FIS2005-00319 }}
\author{Luis Mart\'{\i}nez Alonso$^{1}$ and
 Elena Medina$^{2}$\\\\
\emph{ $^1$Departamento de F\'{\i}sica Te\'{o}rica II, Universidad
Complutense}\\ \emph{E28040 Madrid, Spain} \\
\emph{$^2$Departamento de Matem\'{a}ticas, Universidad de
C\'{a}diz}\\\emph{ E11510, Puerto Real, C\'{a}diz, Spain}}

\date{}
\maketitle
\begin{abstract}
A scheme for solving quasiclassical string equations is developped
to prove that genus-zero Whitham hierarchies describe the
deformations of planar domains determined by rational conformal
maps. This property is applied in  normal matrix models to show that
deformations of simply-connected supports of eigenvalues under
changes of coupling constants are governed by genus-zero Whitham
hierarchies.
\end{abstract}

\vspace*{.5cm}

\begin{center}\begin{minipage}{12cm}

\end{minipage}
\end{center}
\newpage

\section{Introduction}

Conformal mapping methods have been effectively applied in the
analysis of interfacial free-boundary problems involving planar
domains\cite{crowdy}. They have provided many exact solutions
\cite{pol}-\cite{rich} which stimulated the research on possible
underlying integrable structures. Thus, Wiegmann and Zabrodin
discovered \cite{zab1}-\cite{zab2} that deformations of
simply-connected domains with respect to changes of  their exterior
harmonic moments, treated as independent variables, are described by
the dispersionless Toda hierarchy. They also formulated an
algebro-geometric analysis \cite{zab3}-\cite{zab4} of the
deformations of multiply-connected domains in terms of Whitham
equations for Abelian differentials.

 Recent research has shown \cite{cro}-\cite{sha} that many exact solutions of Laplacian growth models correspond to
a special type of domains called \emph{algebraic} or
\emph{quadrature domains}. In the simply-connected case the
complement of a quadrature domain $D$ is the image of the exterior
of the unit disk under a conformal map given by a rational function
\begin{equation}\label{0.1}
z(w)={r}\,w+\sum_{n=0}^{N_0} \frac{u_{0,n}}{w^n}+\sum_{s=1}^k
\sum_{n=1}^{N_s}\frac{{u_{s,n}}}{(w-{a_s})^n},
\end{equation}
where the coefficient $r$ is a positive number and the $k$ poles
$a_s\neq 0$ lie inside the unit circle. In this work we prove that
the deformations of rational conformal maps under changes of the
parameters $(r,u_{0,n},u_{s,m},a_s)$ such that $z(\bar{a}_s^{-1})$
are kept constant, turn out to be described by a solution of the
genus-zero Whitham hierarchy W$(n)$ with $n=2k+2$ punctures
\cite{krich}. It should be noted that according to a recent general
result by Takasaky \cite{tak} the Whitham hierarchy  W$(n)$ is the
quasiclassical limit of the $n$-component KP hierarchy.

Our analysis is based on solving a system of quasiclassical string
equations which leads to the characterization of the conformal map
\eqref{0.1} as a function $z(w,\bt)$ of the Whitham times  $\bt$.
For $k=0$ our result agrees  with \cite{zab1} (see also \cite{lue})
since W$(2)$ is the dispersionless Toda hierarchy. However, for
$k\geq 1$ the analysis of \cite{zab1} does not apply because the
Schwarz function of the boundary of $D$ has poles outside $D$,
consequently there are infinite exterior harmonic moments of $D$
different from zero and, since $z(w)$ depends on a finite number of
parameters only, these harmonic moments are not independent
variables.

Deformations of quadrature domains naturally arise in the analysis of partition functions of $N\times N$ normal matrix models \cite{zab5}
\begin{equation}\label{0.2}\everymath{\displaystyle}
Z_N=\int e^{\frac{1}{\hbar} W(M,M^{\dagger})} \d M\,\d M^{\dagger},
\end{equation}
with \emph{quasiharmonic} potentials $W(z,\bar{z}):=-z\bar{z}+V(z)+\overline{V(z)}$  of the form
\begin{equation}\label{0.3}
V(z)=\sum_{n=1}^{N_0+1}z^n t_{0,n}+\sum_{s=1}^k
\Big(-t_{s,0}\log(z-\beta_s)+\sum_{n=1}^{N_s-1}\frac{t_{s,n}}{(z-\beta_s)^n}\Big).
\end{equation}
Integrating over eigenvalues and ignoring normalization factors, the
partition function reduces to
\begin{equation}\label{0.2a}\everymath{\displaystyle}
Z_N=\int\prod_{i>j}|z_i-z_j|^2\;e^{\frac{1}{\hbar}\sum_j
W(z_j,\bar{z}_j)}\prod_j \d^2 z_j,
\end{equation}
In the large $N$ limit ($N\rightarrow\infty,\;\hbar N \; \mbox{fixed}$) the eigenvalues densely occupy a bounded quadrature domain $D$ in the complex
plane (\emph{the support of eigenvalues}). As a consequence of our analysis we prove  that for simply-connected  supports of eigenvalues the corresponding rational conformal map $z=z(w,\bt)$ as a function of the coupling constants $\bt$ of the partition function
represents a solution of the Whitham hierarchy W$(2k+2)$.

\section{String equations in Whitham hierarchies}
The elements of the phase space for a
genus-zero  Whitham hierarchy W$(M+1)$ are characterized by $M+1$ \emph{punctures}  $q_{\alpha},\; (\alpha=0,\ldots,M)$, where $q_0:=\infty$, of the extended
complex $p$-plane  and an associated set of local coordinates of
the form
\begin{equation}\label{1.1}\everymath{\displaystyle}
      z_0=
        p+\sum_{n=1}^\infty \frac{c_{0,n}}{p^n};\quad
        z_i=\dfrac{d_{i}}{p-q_i}+\sum_{n=0}^\infty d_{i,n} (p-q_i)^n,
       \quad i=1,\dots, M.
\end{equation}
In what follows Greek and Latin suffixes will label
indices of the sets $\{0,\ldots,M\}$ and $\{1,\ldots,M\}$,
respectively. We will henceforth suppose that there exist positively oriented closed
curves
$\Gamma_{\mu}$ in the complex planes of the variables $z_{\mu}$ such that each function $z_{\mu}(p)$
determines a conformal map of the right-exterior of a
circle $\gamma_{\mu}:=z_{\mu}^{-1}(\Gamma_{\mu})$  on the
exterior of $\Gamma_{\mu}$ (we will assume that the circle $\gamma_0$   encircles all the $\gamma_i $)  (see figure 1).

The flows of the Whitham hierarchy can be formulated as the
following infinite system of quasiclassical Lax equations
\begin{equation}\label{wh} \frac{\partial z_{\alpha}}{\partial
t_{\mu n}}=\{\Omega_{\mu n}, z_{\alpha}\},
\end{equation}
associated to the series of time parameters $\{t_{0,n}:n\geq 1\; ; t_{i,n}:i=1,\ldots,M, n\geq 0\}$. Here the Poisson bracket is defined as $\{F,G\}:=\partial_p F\;\partial_x G-\partial_x F\;\partial_pG$, $x:=t_{01}$
and the Hamiltonian functions are
\begin{equation}\label{1.3}\everymath{\displaystyle}
\Omega_{\mu n}:=(z_\mu^n)_{(\mu,+)} ,\; (n\geq 1) ;\quad
\Omega_{i0}:=-\log(p-q_i),
\end{equation}
where $(\cdot)_{(i,+)}$ and $(\cdot)_{(0,+)}$ stand for the
projectors on the subspaces generated by
$\{(p-q_i)^{-n}\}_{n=1}^\infty$ and $\{p^n\}_{n=0}^\infty$ in the
corresponding spaces of Laurent series. These hierarchies \cite{tak}
are the dispersionless limits of the multi-component KP hierarchies.
In particular, for $M=0$ and $M=1$ they represent the dispersionless
versions of the KP and Toda hierarchies, respectively.
\begin{figure}\label{fig 1}
\begin{center}
\psfrag{p}{$p$-plane}\psfrag{z}{$z_\mu$-plane}
\psfrag{a}{$\gamma_0$}\psfrag{b}{$\gamma_1$}\psfrag{c}{$\gamma_2$}
\psfrag{d}{$\gamma_3$} \psfrag{e}{$\Gamma_{\mu}$}
\includegraphics[width=13cm]{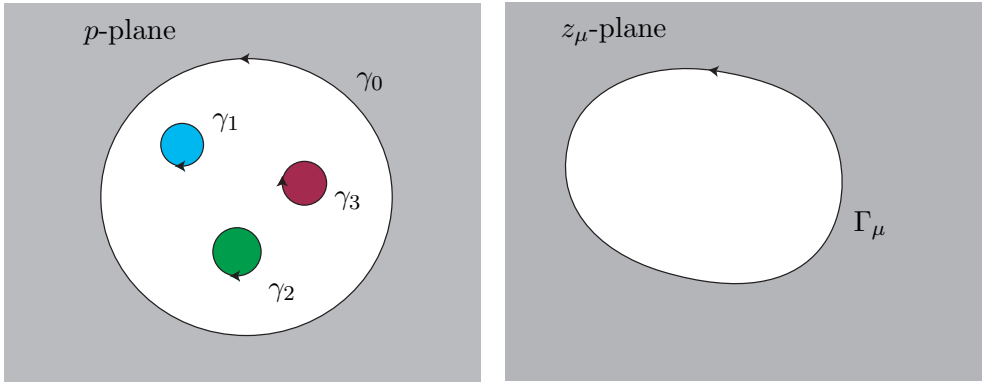}
\end{center}\caption{Right-exteriors of $\gamma_{\mu}$ and $\Gamma_{\mu}$}
\end{figure}

 In our analysis we will use an extended Lax formalism
with  Orlov functions
\begin{equation}\label{1.2}
m_{\alpha}(z_{\alpha},\bt)=\sum_{n=1}^{\infty}nt_{\alpha
n}z_{\alpha}^{n-1}+\frac{t_{\alpha 0}}{z_{\alpha}}+
\sum_{n\geq2}\frac{v_{\alpha n}}{z_{\alpha}^n},
\end{equation}
which verify the same Lax equations \eqref{wh} as the variables
$z_\alpha$, and such that
$
\{z_\alpha,m_\alpha\}=1.
$
The parameter $t_{00}$ in \eqref{1.2} is defined by
\begin{equation}\label{1.2a}
t_{0
0}:=-\sum_{i=1}^Mt_{i 0},
\end{equation}

 The Whitham hierarchy can be formulated as the following system of equations
\begin{equation}\label{2.a}
\d z_{\alpha}\wedge\d m_{\alpha}=\d \omega,\quad \forall \alpha,
\end{equation}
where $\omega$ is the one-form
\begin{equation}\label{2.aaa}
\omega:=\sum_{\mu,n}\Omega_{\mu n} \d t_{\mu n}.
\end{equation}
To see how to get from the system \eqref{2.a} to the  Whitham hierarchy, note that by identifying the coefficients of $\d p\wedge\d t_{\mu
n}$ and $\d x\wedge\d t_{\mu n}$ in \eqref{2.a} we obtain

\begin{equation}\label{2.b}\everymath{\displaystyle}
\frac{\partial z_{\alpha}}{\partial p}\frac{\partial
m_{\alpha}}{\partial t_{\mu n}} -\frac{\partial m_{\alpha}}{\partial
p}\frac{\partial z_{\alpha}}{\partial t_{\mu n}}=\frac{\partial
\Omega_{\mu n}}{\partial p},\quad \quad \frac{\partial
z_{\alpha}}{\partial x}\frac{\partial m_{\alpha}}{\partial t_{\mu
n}} -\frac{\partial m_{\alpha}}{\partial x}\frac{\partial
z_{\alpha}}{\partial t_{\mu n}}=\frac{\partial \Omega_{\mu
n}}{\partial x}.
\end{equation}
and, in particular, since $\Omega_{01}=p$, for $(\mu,n)=(0,1)$,
the system \eqref{2.b} implies
$
\{z_{\alpha},m_{\alpha}\}=1.
$
Thus, using this fact and solving  \eqref{2.b} for
$\partial_{t_{\mu n}} z_{\alpha}$ and
$\partial_{t_{\mu n}} m_{\alpha}$, the Lax equations for $(z_{\alpha},m_{\alpha})$ follow.

A natural form of characterizing solutions of Whitham hierarchies is
provided by systems of string equations
\begin{equation}\label{2.1}
\begin{cases}
P_i(z_i,m_i)=P_0(z_0,m_0),\\
Q_i(z_i,m_i)=Q_0(z_0,m_0),
\end{cases}\quad i=1,2,\dots,M,
\end{equation}
where $\{P_\alpha,Q_\alpha\}_{\alpha=0}^M$ satisfy
$\{P_{\alpha}(p,x),Q_{\alpha}(p,x)\}=1$. Given a solution $(z_{\alpha}(p,\bt),m_{\alpha}(p,\bt))$ of a system \eqref{2.1}, if we denote
\[
\mathcal{P}_{\alpha}(p,\bt):=
P_{\alpha}(z_{\alpha}(p,\bt),m_{\alpha}(p,\bt)),\quad
\mathcal{Q}_{\alpha}(p,\bt):=Q_{\alpha}(z_{\alpha}(p,\bt),m_{\alpha}(p,\bt)),
\]
it is clear that
\begin{equation}\label{2.6}
\d \mathcal{P}_{\alpha}\wedge\d \mathcal{Q}_{\alpha}=\d \mathcal{P}_{\beta}\wedge\d \mathcal{Q}_{\beta},\quad \forall \alpha,\beta.
\end{equation}
On the other hand $\{P_{\alpha}(p,x),Q_{\alpha}(p,x)\}=\{z_{\alpha},m_{\alpha}\}=1$, so that  solutions of a system of string equations verify
\begin{equation}\label{2.8}
\d\mathcal{P}_{\alpha}\wedge\d\mathcal{Q}_{\alpha}=\d z_{\beta}\wedge\d m_{\beta},\quad \forall \alpha,\beta.
\end{equation}
\vspace{0.3truecm}

\noindent
{\bf Theorem}
\emph{Let $(z_{\alpha}(p,\bt),m_{\alpha}(p,\bt))$ be a  solution of \eqref{2.1} which admits  expansions
of the form  \eqref{1.1},\eqref{1.2},\eqref{1.2a} and such that the coefficients of the
the two-forms \eqref{2.8}  are meromorphic functions of the complex variable $p$ with finite poles at
$\{q_1,\dots,q_M\}$ only. Then $(z_{\alpha}(p,\bt),m_{\alpha}(p,\bt))$
is a solution of the Whitham hierarchy.}

\begin{proof}

In view of the hypothesis of the theorem the coefficients of
the two-forms \eqref{2.8} with respect to the basis
\[
\{\d p\wedge\d t_{\alpha n},\quad\d t_{\alpha n}\wedge\d t_{\beta m} \}
\]
are determined by their  principal parts at $q_{\mu},
(\mu=0,\ldots ,M)$, so that by taking \eqref{2.8} into account we may write
\[
\d z_{\alpha}\wedge\d m_{\alpha}=\sum_{\mu=0}^M (\d z_{\mu}\wedge\d m_{\mu})_{(\mu,+)},\quad \forall \alpha.
\]
Moreover the terms in these decompositions  can be found by using the expansions \eqref{1.2} of the functions $m_{\mu}$ as follows
\begin{align*}
& \d z_{\mu}\wedge \d m_{\mu}=\d z_{\mu}\wedge\Big(
\sum_{n=1}^{\infty}nz_{\mu}^{n-1}\d t_{{\mu}n}+\frac{\d t_{{\mu}
0}}{z_\mu}+ \sum_{n\geq2}\frac{\d
v_{{\mu} n}}{z_{\mu}^n}\Big)\\
&=\d \Big(\sum_{n=1}^{\infty}z_{\mu}^{n}\d t_{{\mu}n}+\log z_{\mu}
\d t_{{\mu} 0}- \sum_{n\geq2}\frac{1}{n-1}\frac{\d
v_{{\mu}n}}{z_{\mu}^{n-1}}\Big),
\end{align*}
so that
\[
(\d z_{\mu}\wedge\d m_{\mu})_{(\mu,+)}=\d
\Big(\sum_{n=1}^{\infty}(z_{\mu}^{n})_{(\mu,+)}\d
t_{{\mu}n}-(1-\delta_{\mu 0})\log (p-q_{\mu}) \d
t_{{\mu}0}\Big)
=\d\Big(\sum_{n}\Omega_{\mu n} \d t_{\mu n}\Big).
\]
Thus we find
\[
\d z_{\alpha}\wedge\d m_{\alpha}=\d\omega=\d \Big(\sum_{\mu,n}\Omega_{\mu n} \d t_{\mu n}\Big),\quad \forall \alpha,
\]
and, consequently, this proves that the functions  $(z_{\alpha}(p,\bt),m_{\alpha}(p,\bt))$
determine a solution of the Whitham hierarchy.
\end{proof}

\section{Integrable dynamics of quadrature domains}

Let us consider a rational conformal map
 $z(w)$ of the form \eqref{0.1} with  $k$ poles $a_s$ inside the unit circle and define
\begin{equation}\label{0.3a}
\tilde{z}(w):=\overline{z(\bar{w}^{-1})}=\frac{r}{w}+\sum_{n=0}^{N_0}
\bar{u}_{0,n}\,w^n+\sum_{s=1}^k
\sum_{n=1}^{N_s}\frac{\bar{u}_{s,n}\,w^n}{(1-w\bar{a}_s)^n}.
\end{equation}
In order to establish the connection between the deformations of the
conformal map $z(w)$ and the genus-zero Whitham hierarchies we introduce the change of variable
\[
 p=R\,w:=r\,w+u_{00},
\]
where $r$ and $u_{00}$ are the first coefficients of $z(w)$ in
\eqref{0.1}. As a function of the new variable $p$ the conformal map
is normalized at infinity
\[
z(p)=p+\mathcal{O}(1/p),\quad p\rightarrow\infty.
\]
Moreover, if we define
\[
q_0:=\infty,\quad q_s:=R\,b_{s},\quad q_{s+k}:=R\,a_{s},\quad
q_{2k+1}:=u_{00},
\]
it is clear that $z(p)$ and $\tilde{z}(p)$ become rational functions
of $p$ with poles at $(q_0,q_s,q_{2k+1})$ and
$(q_0,q_{s+k},q_{2k+1})$ respectively.

We are going to prove that deformations of $z(w)$ with respect to
the coefficients
\[
\bu:=(r,u_{0,n},u_{s,m},a_s),
\]
such that $\beta_s=z(w)|_{b_s}$ are kept constant, are described by
the Whitham hierarchy  W$(2k+2)$. To this end we introduce Whitham
variables $(z_{\alpha},m_{\alpha})$ on the $2k+2$ punctures $q_{\alpha}$ of the $p$-plane

\begin{equation}\label{3.1}
\everymath{\displaystyle}
\begin{cases}
z_0=z,\quad m_0=\tilde{z},\quad (\mbox{near $q_0=\infty$})\\\\
z_{s}=\frac{1}{z-\beta_s},\quad m_{s}=-(z-\beta_s)^2\,\tilde{z}
,\quad (\mbox{near $q_{s}$})\\\\
z_{s+k}=\frac{1}{\tilde{z}-\bar{\beta_s}},\quad
m_{s+k}=(\tilde{z}-\bar{\beta_s})^2\,{z} \quad (\mbox{near $q_{s+k}$}).\\\\
z_{2k+1}=\tilde{z},\quad m_{2k+1}=-z,\quad (\mbox{near $q_{2k+1}$})
\end{cases}
\end{equation}
It is clear that these variables are rational functions of $p$ with possible poles at the punctures $q_{\alpha}$ only. Moreover, they satisfy the system of string equations

\begin{equation}\label{3.2}
\everymath{\displaystyle}
\begin{cases}
\frac{1}{z_{s}}+\beta_s=m_{s+k}z_{s+k}^2 =-m_{2k+1}=z_0
\\\\
-m_{s}z_{s}^2=\frac{1}{z_{s+k}}+\bar{\beta_s}=z_{2k+1}=m_0.
\end{cases}
\end{equation}
Obviously the functions $z_{\alpha}$ are of the form \eqref{1.1}. On the other hand due to \eqref{0.1} and \eqref{0.3a} it follows that the functions $m_{\alpha}$ defined in \eqref{3.1} verify expansions of the form \eqref{1.2}
\begin{equation}\label{3.2b}\everymath{\displaystyle}
\begin{array}{ll}
m_0=\tilde{z}=\sum_{n=1}^{N_0+1}nt_{0n}z_0^{n-1}+\frac{t_{00}}{z_0}+\cdots& w\rightarrow\infty,\\\\
m_s=-z_s^{-2}\tilde{z}=\sum_{n=1}^{N_s-1}nt_{sn}z_s^{n-1}+\frac{t_{s0}}{z_s}+\cdots& w\rightarrow b_s,\\\\
m_{s+k}=z_{s+k}^{-2} z=\sum_{n=1}^{N_s-1}nt_{s+kn}z_{s+k}^{n-1}+\frac{t_{s+k0}}{z_{s+k}}+\cdots& w\rightarrow a_s,\\\\
m_{2k+1}=-z=\sum_{n=1}^{N_0+1}nt_{2k+1n}z_{2k+1}^{n-1}+\frac{t_{2k+10}}{z_{2k+1}}+\cdots& w\rightarrow 0,
\end{array}
\end{equation}
where the time parameters $\bt:=(t_{\alpha,n})$
are rational functions in $\bu$
\begin{equation}\label{3.2c}
t_{\alpha n}=Q_{\alpha n}(\bu).
\end{equation}

The functions $Q_{\alpha n}(\bu)$  satisfy certain constraints
which can be characterized by considering the map
\begin{equation}\label{c}
C:f\mapsto Cf,\quad Cf(w):=\overline{f(\bar{w}^{-1})}.
\end{equation}
Observe that in terms of the variable $p$
\begin{equation}\label{cc}
Cf(p)=\overline{f(\mathcal{I}p)},
\end{equation}
where $\mathcal{I}p=r^2/(\overline{p-u_{00}})+u_{00}$ is the
inversion with respect to the circle $|p-u_{00}|^2=r^2$. From
\eqref{3.1} it is clear that
\begin{equation}\label{3.3}
z_{2k+1}=Cz_0,\quad m_{2k+1}=-Cm_0,\quad z_{s+k}=Cz_s, \quad
m_{s+k}=-Cm_s,
\end{equation}
which implies
\begin{equation}\label{3.4}
Q_{2k+1,n}(\bu)=-\overline{Q_{0,n}(\bu)},\quad Q_{s+k,n}=-\overline{Q_{s,n}(\bu)}.
\end{equation}
Furthermore, we can prove that
\begin{equation}\label{3.4a}
\sum_{\alpha}Q_{\alpha 0}(\bu)=0.
\end{equation}
Indeed
from \eqref{3.1} we deduce
\[
m_0\,\d z_0=m_{s}\,\d z_{s}=\tilde{z}\,\d z,\quad m_{2k+1}\d
z_{2k+1}=m_{s+k}\d z_{s+k}=-z\,\d\tilde{z}.
\]
Hence
\[\everymath{\displaystyle}
2\pi i\sum_{\alpha}
Q_{\alpha 0}=\sum_{\alpha}\oint_{\Gamma_{\alpha}}m_{\alpha}\d
z_{\alpha}=\sum_{\alpha}\oint_{\gamma_{\alpha}}\tilde{z}\,\partial_p
z\d p=0,
\]
where we have taken into account that $\tilde{z}\,\partial_p
z$ is a rational function of $p$ with poles at the punctures $q_{\alpha}$ only, and
the fact that
\[
\sum_{\alpha}\gamma_{\alpha}\sim 0 \quad \mbox{in}\quad
\mathbb{C}\setminus\{q_1,\dots,q_{2k+1}\}.
\]
Notice that due to \eqref{3.4} the constraint \eqref{3.4a} can be rewritten as
\begin{equation}\label{3.4b}
Im\,Q_{00}+\sum_{s}Im \,Q_{s 0}(\bu)=0.
\end{equation}
Under appropriate conditions one can determine $\bu$  as a function of $(\bt,\beta_s,\bar{\beta}_s)$. To this end we consider the system
\begin{equation}\label{3.5}
\begin{cases}
t_{\alpha n}=Q_{\alpha n}(\bu),\\\\
\beta_s=z(b_s,\bu),
\end{cases}
\end{equation}
where the time parameters
\begin{equation}\label{3.6}
\bt:=(t_{\alpha,n}),\quad\alpha=0,\ldots, 2k+1; n=0,\ldots
\widetilde{N}_{\alpha};\;\;
\widetilde{N}_0=\widetilde{N}_{2k+1}=N_0+1,\;
\widetilde{N}_{s}=\widetilde{N}_{s+k}=N_s-1,
\end{equation}
are assumed to satisfy
\begin{equation}\label{3.7}
t_{2k+1,n}=-\overline{t_{0,n}},\quad t_{s+k,n}=-\overline{t_{s,n}},
\quad \sum_{\alpha}t_{\alpha 0}=0.
\end{equation}
Firstly, we observe that $\bu$ constitutes a set of
\begin{equation}\label{3.8}
2(N_0+\sum_s N_s+k)+3,
\end{equation}
real variables given by $r$ and the real and imaginary parts of
$(u_{0,n},u_{s,m},a_i)$.
On the other hand, in view of \eqref{3.4} and \eqref{3.4a} we may
ignore the equations corresponding to $\alpha=s+k, 2k+1$ in \eqref{3.5}. In this way the system \eqref{3.5} reduces to $2(N_0+2+\sum_s N_s+k)$ real equations, but due to \eqref{3.4b} one
of them is a consequence of the others. Therefore, we are lead to a system of equal number of equations and unknowns
  which under appropriate
conditions will determine $\bu$, and consequently
$(z_{\alpha},m_{\alpha})$ as functions of
$(\bt,\beta_s)$.

In this way we have determined a rational solution
$(z_{\alpha}(p),m_{\alpha}(p))$ of the system of string equations
\eqref{3.2} which depends on $(\bt,\beta_1,\ldots,\beta_k)$ and
satisfies the asymptotic conditions \eqref{1.1} and \eqref{1.2}.
Therefore, from the above Theorem  we conclude that this  solution
evolves with respect to $\bt$ according to the Whitham hierar
chy
W$(2k+2)$.

It is interesting to notice the following identity involving the times $t_{\alpha 0}$ and the area $T$ of the domain $D$. Let $\Gamma$ be the positively oriented boundary of $D$ and take small positively oriented closed curves
$\Gamma'_s$ around the points $\beta_s$ of the $z$-plane, then we have that
\begin{align}\label{are}
\everymath{\displaystyle}
\nonumber t_{00}&=\frac{1}{2\pi i}\oint_{\Gamma_0}\tilde{z}\d z=
\sum_s\frac{1}{2\pi i}\oint_{\Gamma'_s}\tilde{z}\d z
+\frac{1}{2\pi i}\oint_{\Gamma}\tilde{z}\d z\\\\
\nonumber &=\sum_s\frac{1}{2\pi i}\oint_{\Gamma_s}z_s^{-2}\tilde{z}\d z_s
+\frac{1}{2\pi i}\oint_{\Gamma}\bar{z}\d z
=-\sum_s t_{s0}+\frac{1}{\pi}T.
\end{align}

\section{Exact solutions}

The above analysis can be used to generate solutions of W$(2k+2)$ of the form \eqref{0.1}. Let us consider the case in which only simple poles arise
\[
 z=rw+u_0+\sum_{s=1}^k\frac{v_s}{w-a_s},\quad
\tilde{z}=\frac{r}{w}+\bar{u}_0+\sum_{s=1}^k\frac{\overline{v}_sw}{1-\overline{a}_s
w}.
\]
By identifying the coefficients of $z_0^0$ and  $z_0^{-1}$ in the expansion of $m_0=\tilde{z}$ as $z_0=z\rightarrow\infty$
($w\rightarrow\infty$), we
get

\begin{equation}\label{5.1}
\bar{u}_0 - \sum_{s=1}^k\frac{\overline{v}_s}{\overline{a}_s}
=t_{0,1},\quad
r\left(r-\sum_{s=1}^k\frac{\overline{v}_s}{{\overline{a}_s}^2}\right)=t_{0,0},
\end{equation}
and identifying the coefficient of $z_s^{-1}$ in the expansion of $m_s=-(z-\beta_s)^2\tilde{z}$ as
$z_s\rightarrow\infty$
($w\rightarrow b_s$) yields

\begin{equation}\label{5.2}
\frac{\bar{v}_s}{\bar{a}_s^2}\left(r-\sum_{s'=1}^k\frac{v_{s'}\bar{a}_s^2}{(1-a_{s'}\bar{a}_{s})^2}\right)=t_{s,0}.
\end{equation}
Finally, the equations $z(w)|_{w=b_s}=\beta_s$ read
\begin{equation}\label{5.3}
\frac{r}{\bar{a}_s} + u_0 +\sum_{s'=1}^k \frac{v_{s'}\bar{a}_s}{1 - \bar{a}_sa_{s'}}=\beta_s.
\end{equation}
The system \eqref{5.1}-\eqref{5.3} determines $(r,u_0,v_s,a_s)$ in terms of $(t_{0,0},t_{0,1},t_{s,0},\beta_s)$.

\vspace{0.3cm}
\noindent
\emph{Examples}
\vspace{0.3cm}

A solution of W$(4)$ is obtained by deforming the conformal map (aircraft wind)
$$z=r\,w+u_0+\frac{v}{w-a},\qquad
\tilde{z}=\frac{r}{w}+\bar{u}_0+\frac{\bar{v}w}{1-\bar{a}w}.$$

\vspace{5mm}

\begin{figure}[h]
\centering
\includegraphics[width=8cm]{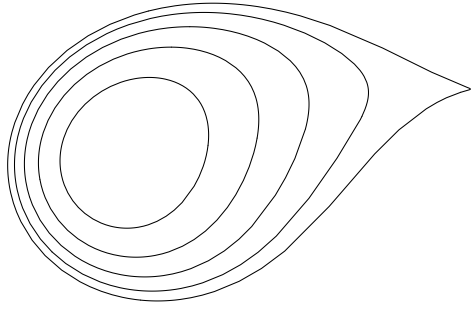}
\caption{Solution corresponding to $k=1$}
\end{figure}

\vspace{5mm}

The solution of the corresponding equations \eqref{5.1}-\eqref{5.3} is given by:
$$a\,=\,\frac{r^2 -At_{0,0}}
        {r\,(\bar{\beta}-t_{0,1})(1 -A ) },\quad
v\,=\,\frac{(r^2 - \overline{t}_{0,0})\,(r^2 -t_{0,0}A)^2}
{r^3\,(\bar{\beta}-t_{0,1})^2\,( 1- A)^2},
          $$
$$u_0\,=\,\frac{r^4 -r^2\,t_{0,0}A +r^2\,\overline{t}_{0,1}\,(\bar{\beta}-t_{0,1})(1-A) +
          |t_{0,0}|^2A -r^2\,\overline{t}_{0,0}}
          {r^2\,(\bar{\beta}-t_{0,1})(1-A)},$$
with
$$r^2=\frac{2\,|t_{0,0}|^2\,A^2}
{|\bar{\beta}-t_{0,1}|^2\,(2\,A^3-3\,A^2+A)+(t_{0,0}+t_{1,0})(A-1) +
(t_{0,0}+\overline{t}_{0,0})\,A},$$ where $A=|a|^2$ is implicitly
defined by
$$\everymath{\displaystyle}\begin{array}{l}
2|\bar{\beta}-t_{0,1}|^4A^5+\Big[2(t_{0,0}+t_{1,0})-2(t_{0,0}+\overline{t}_{0,0})-5|\bar{\beta}-t_{0,1}|^2\Big]
|\bar{\beta}-t_{0,1}|^2A^4\\  \\
+4\Big[(t_{0,0}+\overline{t}_{0,0})-(t_{0,0}+t_{1,0})+|\bar{\beta}-t_{0,1}|^2\Big]|\bar{\beta}-t_{0,1}|^2A^3\\  \\
+\Big[(t_{0,0}+t_{1,0})^2-(t_{0,0}+\overline{t}_{0,0})^2+4|t_{0,0}|^2-2(t_{0,0}+\overline{t}_{0,0})
|\bar{\beta}-t_{0,1}|^2\\  \\
+2(t_{0,0}+t_{1,0})|\bar{\beta}-t_{0,1}|^2-|\bar{\beta}-t_{0,1}|^4\Big]A^2
-2(t_{0,0}+t_{1,0})^2A+(t_{0,0}+t_{1,0})^2=0.
\end{array}$$

It can be proved that a  reduction
of the system \eqref{5.1}-\eqref{5.3} is obtained by setting
\begin{equation}\label{5.4}
\everymath{\displaystyle}
z=rw+u+\sum_{s=1}^l\big(\frac{v_s}{w-a_s}+\frac{\bar{v}_s}{w-\bar{a}_s}\Big),\quad
\tilde{z}=\frac{r}{w}+u+\sum_{s=1}^l\Big(\frac{\bar{v}_sw}{1-\bar{a}_sw}+\frac{v_sw}{1-a_sw}
\Big),
\end{equation}
and by assuming
\[u,\,t_{00},\,t_{01}\in \mathbb{R},\; t_{s+l,0}=\bar{t}_{s,0},\;
\beta_{s+l}=\bar{\beta}_s.
\]
 In the simplest case $l=1$ the reduced system of implicit equations reads
\begin{equation}\label{5.5}\everymath{\displaystyle}\begin{array}{l}
u- \frac{v_1}{a_1} -\frac{\bar{v}_1}{\bar{a}_1} =t_{0,1},\quad
r\left(r-\frac{v_1}{{a_1}^2}-\frac{\bar{v}_1}{{\bar{a}_1}^2}\right)=t_{0,0},\\  \\
\bar{v}_1\left(\frac{r}{\bar{a}_1^2}-\frac{\bar{v}_1}{(1- \bar{a}_1^2)^2}- \frac{v_1}{(1 - a_1\bar{a}_1)^2}\right)=t_{1,0},\\  \\
\frac{r}{\bar{a}_1} + u + \frac{v_1\bar{a}_1}{1 - a_1\bar{a}_1}
+\frac{\bar{v}_1\bar{a}_1}{1 -\bar{a}_1^2}=\beta_1,
\end{array}
\end{equation}
and determines a solution of W$(6)$.

Figures 2 and 3 exhibit  deformations of domains with respect to
changes of the area $T$ such that the Whitham times $t_{s0}$ are
kept constant. Observe that the evolution of the boundary develops
cusp-like singularities \cite{ben}.

\vspace{5mm}

\begin{figure}
\centering
\includegraphics[width=6cm]{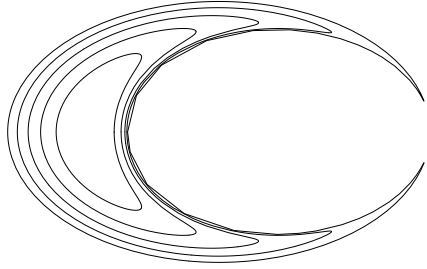}
\caption{Solution corresponding to $k=2$}
\end{figure}

\vspace{5mm}

\section{Whitham times and coupling constants of normal matrix models}

Let us assume that the support of eigenvalues  of the normal matrix model \eqref{0.2} is a simply-connected domain $D$. For example this is  the case if $k=1$ \cite{zab4}. Then by applying the saddle point method to the large $N$ limit of  \eqref{0.2a} it follows that \begin{equation}\label{4.1}
\everymath{\displaystyle}
\bar{z}=V'(z)+\frac{1}{2\pi i}\int\!\!\!\int_D \frac{\d z'\wedge\d\bar{z}'}{z'-z},\quad z\in D.
\end{equation}
Let us consider now the rational conformal map $z(w)$ associated to
$D$, from \eqref{4.1} it follows that the function $\tilde{z}(w)$
can be extended as a meromorphic function of $z$ outside $D$ by
\begin{equation}\label{4.2}
\everymath{\displaystyle}
\tilde{z}(z)=V'(z)+\frac{1}{2\pi i}\int\!\!\!\int_D \frac{\d z'\wedge\d\bar{z}'}{z'-z}.
\end{equation}
In fact $\tilde{z}(z)$ represents the Schwarz function of the boundary $\Gamma$ of $D$. By using \eqref{0.3} and \eqref{3.1} one can rewrite \eqref{4.2} as
\begin{equation}\label{4.3}
\everymath{\displaystyle}
\tilde{z}(z)=\sum_{n=1}^{N_0+1}nz_0^{n-1}t_{0n}-
\sum_{s=1}^k(\sum_{n=1}^{N_s-1}nz_s^{n+1}t_{sn}+z_st_{s0})+
\frac{1}{2\pi i}\int\!\!\!\int_D \frac{\d z'\wedge\d\bar{z}'}{z'-z}.
\end{equation}
By comparing this identity with our definition \eqref{3.2b} of the Whitham times  and by taking into account \eqref{are},
we conclude that the Whitham times coincide with the coupling constants of the normal matrix model.

Finally, we notice that by using the strategy deployed in the proof of the Theorem of Section 2 and by taking \eqref{4.3} into account, it follows that the deformations with respect to the parameters $\beta_s$ and
$\bar{\beta}_s$ are described by the flows
\begin{equation}\label{whe}
\frac{\partial z_{\alpha}}{\partial \beta_s}=t_{s,0}\frac{\partial
z_{\alpha}}{\partial
t_{s,1}}+\sum_{n=1}^{N_s-2}nt_{s,n}\frac{\partial
z_{\alpha}}{\partial t_{s,n+1}}+(N_s-1)t_{s,N_s-1}
\{(z_{s}^{N_s})_{(s,+)}, z_{\alpha}\},
\end{equation}
and
\begin{equation}\label{whee}
\frac{\partial z_{\alpha}}{\partial
\bar{\beta}_s}=t_{s+k,0}\frac{\partial z_{\alpha}}{\partial
t_{s+k,1}}+\sum_{n=1}^{N_s-2}nt_{s+k,n}\frac{\partial
z_{\alpha}}{\partial t_{s+k,n+1}}+(N_s-1)t_{s+k,N_s-1}
\{(z_{s+k}^{N_s})_{(s+k,+)}, z_{\alpha}\}.
\end{equation}

\end{document}